\documentclass{jps-cp}

\usepackage{amsmath}
\usepackage{amssymb}
\usepackage{amsfonts}
\usepackage{amsthm}
\usepackage{mathtools} 
\usepackage{bm}        
\usepackage{bbm}       
\usepackage{slashed}
\usepackage{cancel}
\usepackage{cases}

\usepackage{graphicx}  
\usepackage{xcolor}

\usepackage{physics}

\DeclareMathOperator{\sn}{sn}
\DeclareMathOperator{\cn}{cn}
\DeclareMathOperator{\dn}{dn}

\title{Thermodynamic Properties of the $S=1/2$ Zigzag Spin Ladder}

\author{Nobuya \textsc{Maeshima}$^{1,2}$, Kouichi \textsc{Okunishi}$^{3,4}$, and Toshiya \textsc{Hikihara}$^{5}$}
\inst{$^{1}$Center for Computational Sciences, University of Tsukuba, Tsukuba, Ibaraki 305-8577, Japan \\
$^{2}$Institute of Pure and Applied Sciences, University of Tsukuba, Tsukuba, Ibaraki 305-8573, Japan \\
$^{3}$Department of Physics, Graduate School of Science, Osaka Metropolitan University, Osaka 558-8585, Japan \\
$^{4}$Nambu Yoichiro Institute of Theoretical and Experimental Physics (NITEP), Osaka Metropolitan University, Osaka 558-8585, Japan \\
$^{5}$Graduate School of Science and Technology, Gunma University, Kiryu, Gunma 376-8515, Japan}

\email{maeshima@ims.tsukuba.ac.jp}

\abst{
We investigate finite-temperature properties of the spin-1/2 zigzag Heisenberg ladder with an anisotropic off-diagonal $\Gamma$ interaction.
Thermal states are calculated by an infinite time-evolving block decimation method combined with purification, while exact diagonalization is used to analyze low-lying excitations.
The entropy is strongly suppressed at low temperatures inside the gapped phases, whereas it remains enhanced near several ground-state phase boundaries.
For the N\'eel phase at $J_2/J_1=0$, the model can be mapped by spin rotations onto the exactly solvable XYZ chain. 
The thermal gap extracted from the low-temperature specific heat agrees with the exact one-spinon gap, while the finite-size excitation gap obtained by exact diagonalization is consistent with a two-spinon excitation.
}

\kword{quantum spin systems, frustration, finite temperature, iTEBD, spinon}

\begin{document}
\maketitle

\section{Introduction}


Frustrated one-dimensional quantum spin systems provide an ideal platform where competing interactions generate a variety of quantum phases. A representative example is the spin-$1/2$ zigzag ladder with antiferromagnetic nearest-neighbor and next-nearest-neighbor exchanges, $J_1$ and $J_2$. The competition between the two exchanges induces a transition between the Tomonaga-Luttinger liquid (TLL) and spontaneously dimerized phases\cite{Haldane1982,OkamotoN1992,NomuraO1994}.
Since various magnetic compounds have been discussed as realizations or approximate realizations of the zigzag ladders\cite{RieraDobry1995,Kikuchi2000,Maeshima2003,Volkova2007,Nilsen2008,Kasinathan2013,Kikuchi2014,Lebernegg2017,Otsuka2018,Fujihala2022}, finite-temperature calculations are important for establishing direct connections between microscopic models and experiments.

Among such zigzag materials, YbCuS$_2$ has recently attracted particular attention because of rapid experimental progress on its magnetic and low-energy properties\cite{OhmagariY2020,HoriF2023,HoriF2025}.
Motivated by these observations, an effective spin-$1/2$ zigzag ladder model with Heisenberg and anisotropic off-diagonal $\Gamma$ interactions has been proposed\cite{SaitoNakaiHotta2024}.
The Hamiltonian is given by
\begin{align}
\mathcal{H}
= \sum_{\rho=1,2}\sum_j \bigl[&J_\rho\,\bm s_j\cdot\bm s_{j+\rho}
+\Gamma_\rho\bigl(s_j^xs_{j+\rho}^y+s_j^ys_{j+\rho}^x\bigr)\bigr],
\label{eq:Hgamma}
\end{align}
where $\bm s_j$ is a spin-$1/2$ operator at the $j$-th site.
The ground-state phase diagram with the parameter $\gamma$ defined by $\Gamma_1=\gamma J_1$ and $\Gamma_2=\gamma J_2$
has been studied in detail and contains several gapped phases, in addition to the gapless TLL phase~\cite{SaitoH2024PRL,SaitoH2024PRB,Hikihara2026}.

In this article, we focus on the finite-temperature properties of this model.
We first describe our computational method, infinite time-evolving block decimation (iTEBD) combined with purification \cite{Vidal2007,FeiguinWhite2005}.
We then discuss the low-temperature thermodynamics of the gapped phases, with particular emphasis on the N\'eel phase along the $J_2/J_1=0$ line.
We show that the thermal gap extracted from the low-temperature specific heat agrees well with the exact one-spinon gap of the XYZ chain.

\section{Method}

Here, we explain our numerical method, iTEBD combined with purification.
In this scheme, the thermal state at temperature $T$ is represented by a purified state $|\Psi(\beta=1/T)\rangle$ in an enlarged Hilbert space consisting of physical and ancilla spins\cite{FeiguinWhite2005}.
At infinite temperature $(\beta=0)$, the purified state is constructed so that each physical spin is maximally entangled with its corresponding ancilla spin.
Starting from the infinite-temperature state $|\Psi(0)\rangle$, the finite-temperature state is obtained as
\begin{equation}
|\Psi(\beta)\rangle=e^{-\beta\mathcal H/2}|\Psi(0)\rangle,
\end{equation}
and the imaginary-time evolution is carried out using iTEBD\cite{Vidal2007} with the second-order Suzuki--Trotter decomposition\cite{Trotter1959,Suzuki1990}.
In the calculations shown below, the bond dimension is increased up to $\chi=200$ when necessary.

Using this method, we evaluate the free energy $F$ and internal energy $E$.
The free energy $F$ is obtained from the dominant eigenvalue of the spatial transfer matrix\cite{OrusVidal2008}
and the associated normalization factors accumulated during the imaginary-time evolution.
The internal energy $E$ is obtained from the expectation value of the local Hamiltonian.
The entropy is then calculated as
$S(T)=\left[ E(T)-F(T)\right]/T$.
All thermodynamic quantities discussed here are evaluated per site in the thermodynamic limit.
Additionally, the specific heat $C(T)=dE/dT$ is obtained by numerical differentiation.
We also perform exact diagonalization (ED) for finite periodic systems to analyze the low-energy excitations contributing to the low-temperature thermodynamics, using the QuSpin\cite{QuSpin2017} and XDiag\cite{xdiag,xdiag-0.4} packages.

\section{Finite-Temperature Properties}

\subsection{Phase diagram and entropy}

\begin{figure}
\centering
\includegraphics[width=115mm]{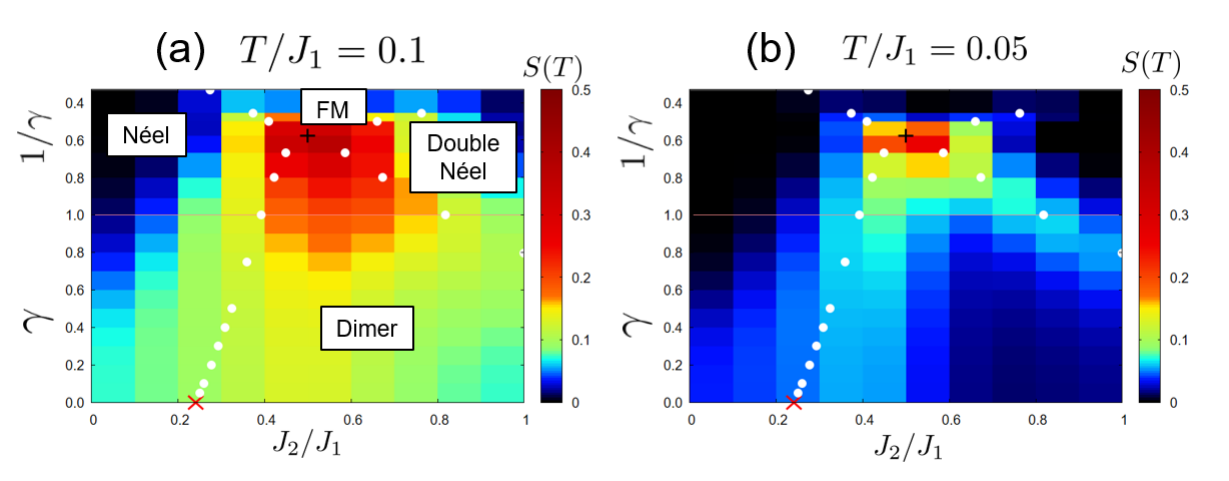}
\caption{Entropy $S(T)$ of the model in Eq.~(\ref{eq:Hgamma})
in the $J_2/J_1$--$\gamma$ parameter plane at (a) $T/J_1=0.1$ and (b) $T/J_1=0.05$.
The red horizontal line indicates $\gamma=1$, and the vertical axis is shown as $1/\gamma$ for $\gamma>1$.
White circles show the ground-state phase boundaries obtained in Ref.~\cite{Hikihara2026}.
The red cross at $(J_2/J_1,\gamma)=(0.2411\ldots, 0)$ marks the transition point between the TLL and the dimer phases on the $\gamma=0$ line~\cite{OkamotoN1992,NomuraO1994},
while the black plus at $(J_2/J_1,\gamma)=(1/2, \sqrt{3})$ marks the multicritical point\cite{SaitoH2024PRL}.
}
\label{fig:ent}
\end{figure}

Figure~\ref{fig:ent} shows the entropy $S(T)$ of the zigzag Heisenberg--$\Gamma$ ladder in Eq.~(\ref{eq:Hgamma})
in the $J_2/J_1$--$\gamma$ parameter plane.
It has been shown that the ground-state phase diagram of the Heisenberg--$\Gamma$ zigzag ladder contains the gapped N\'{e}el (N), dimer (D), double-N\'{e}el (DN), and ferromagnetic (FM) phases, as well as a gapless TLL phase~\cite{SaitoH2024PRL,SaitoH2024PRB,Hikihara2026}.
Comparison with the ground-state phase diagram of Ref.~\cite{Hikihara2026} reveals a clear correspondence between the ground-state phase structure and the low-temperature entropy distribution.
Inside the gapped phases, $S(T)$ is strongly suppressed at low temperatures, reflecting the presence of a finite excitation gap.
In contrast, $S(T)$ remains relatively large near the phase boundaries even at the lower temperature shown here.
This enhancement is particularly pronounced near the N--D and D--DN boundaries, 
suggesting the presence of low-energy excitations in these regions.
A detailed characterization of these excitations is left for future work.

\subsection{Gapped N\'{e}el phase at $J_2/J_1=0$}

In the rest of this article, we focus on the N\'eel phase along the $J_2/J_1=0$ line.
This limit is particularly useful because the exact solution is available~\cite{JohnsonJD1973,Baxter1972,Baxter1982,takahashi1973,takahashi1991}, allowing us to directly compare the thermal gap with the exact one-spinon gap.
The details of the one-spinon gap are summarized in the Appendix.

Under the spin rotations, the Hamiltonian in Eq.~(\ref{eq:Hgamma}) with $J_2=\Gamma_2=0$ is transformed into the XYZ chain\cite{SaitoH2024PRL,SaitoH2024PRB,Hikihara2026}
\begin{equation}
\mathcal H=\sum_j\left[\widetilde{J}_x \widetilde{S}_j^x \widetilde{S}_{j+1}^x +
\widetilde{J}_y \widetilde{S}_j^y \widetilde{S}_{j+1}^y +\widetilde{J}_z \widetilde{S}_j^z \widetilde{S}_{j+1}^z\right],
\end{equation}
where $\widetilde{J}_x=J_1(1-\gamma)$, $\widetilde{J}_y=J_1$, $\widetilde{J}_z=J_1(1+\gamma)$, and $\widetilde{S}_j$ is the rotated spin operator.
The ground state is twofold degenerate, corresponding to the two N\'eel states.

The low-energy excitations are domain walls, i.e., spinons, and the exact one-spinon dispersion\cite{JohnsonJD1973} can be written as
\begin{equation}
\epsilon_{\rm 1s}(q) = A\sqrt{1-k_1^2\cos^2 q},
\label{eq:1spinon}
\end{equation}
where $q$ is the spinon momentum, and the prefactor $A$ and the elliptic parameter $k_1$ are defined in the Appendix.

At low temperatures, the thermodynamics is controlled by one-spinon excitations.
From Eq.~(\ref{eq:1spinon}), the one-spinon dispersion near its minimum at $q=\pi$ can be expanded as
$\epsilon_{\rm 1s}(q)=\Delta_{1\mathrm{s}}+c(q-\pi)^2+\cdots$,
where $c$ is a positive constant and $\Delta_{1\mathrm{s}} = \epsilon_{\rm 1s}(\pi)$ is the one-spinon gap.
Then, the low-temperature specific heat is expected to show the activated behavior~\cite{Troyer1994}
\begin{equation}
C(T)\propto T^{-3/2}\exp\left(-\Delta_{1\mathrm{s}}/T\right).
\label{eq:Cactivated}
\end{equation}
The activation gap in this expression coincides with the soliton gap governing the low-temperature thermodynamics of the XYZ chain obtained by Takahashi~\cite{takahashi1973,takahashi1991}.
In the following, we define the thermal gap $\Delta_{\rm th}$ as the gap parameter extracted from the low-temperature fit to Eq.~(\ref{eq:Cactivated}).

\begin{figure}
\centering
\includegraphics[width=110mm]{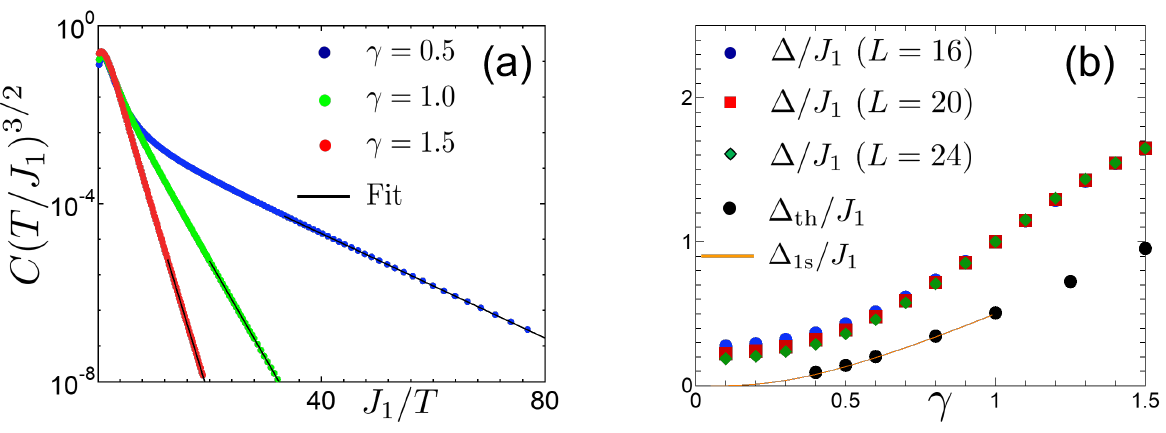}
\caption{
(a) Arrhenius plot of $C(T)T^{3/2}$.
Symbols show the iTEBD data and the solid lines represent the fits to the activated form in Eq.~(\ref{eq:Cactivated}).
(b) Comparison of the thermal gap $\Delta_{\rm th}$ obtained from the fit with the exact one-spinon gap  $\Delta_{1\mathrm{s}}$ and the gap $\Delta$ obtained by the ED calculations for $L=16$, 20, and 24.
}
\label{fig:sgap}
\end{figure}

In Fig.~\ref{fig:sgap}(a),
we fit the low-temperature data of the specific heat $C(T)$ to Eq.~(\ref{eq:Cactivated}).
The resulting thermal gap $\Delta_{\rm th}$  agrees well with the exact one-spinon gap $\Delta_{1\mathrm{s}}$, as shown in Fig.~\ref{fig:sgap}(b).
Consistent with this picture, the excitation gap $\Delta$ obtained by finite-size ED is approximately equal to $2\Delta_{1\mathrm{s}}$,
because the lowest excitation in a periodic system contains two spinons.
This relation between the excitation gap $\Delta$ of the periodic system and the thermal gap $\Delta_{\rm th}$ is familiar from the one-dimensional Ising model\cite{Nagaosa1999}.

For small $\gamma$, the one-spinon gap is exponentially small, and its asymptotic form is derived in the Appendix as
\begin{equation}
\Delta_{1\mathrm{s}}(\gamma) \simeq 2\pi J_1 (1+\gamma)\exp\left(-a/\sqrt\gamma\right),
\end{equation}
where $a=2.912\ldots$ is a constant. 
Thus, the gapless TLL phase exists only at $\gamma=0$, whereas for  finite $\gamma>0$ the system is in the gapped N\'eel phase with a small but finite gap $\Delta_{1\mathrm{s}}$, which is consistent with the ground-state phase diagram of Ref.~\cite{Hikihara2026}.

\section{Summary}

We studied finite-temperature properties of the spin-$1/2$ Heisenberg--$\Gamma$ zigzag ladder using purification-based iTEBD, supplemented by exact diagonalization of low-energy spectra. The entropy is suppressed inside the gapped phases, whereas an enhanced low-temperature entropy remains near the N\'eel--dimer and dimer--double-N\'eel boundaries. In the N\'eel phase at $J_2/J_1=0$, the low-temperature thermodynamics is governed by domain-wall spinons. The thermal gap obtained from the activated specific heat agrees with the exact one-spinon gap of the XYZ chain, while the finite-size ED gap is consistent with a two-spinon excitation.

\section{Acknowledgment}

This work is partially supported by KAKENHI Grant Numbers JP21H05182, JP21H05191, JP24K06881, JP24K06913, and JST CREST Grant Number JPMJCR24I1.
Numerical calculations were carried out on the Pegasus System at the Center for Computational Sciences, University of Tsukuba, and the Miyabi System in the Joint Center for Advanced High Performance Computing.

\appendix
\section{Exact one-spinon gap and small-$\gamma$ asymptotics of the XYZ chain}
\label{app:xyzgap}

Here, we briefly review the exact one-spinon dispersion and one-spinon gap\cite{JohnsonJD1973} used in the main text.
Under the spin rotation $U_z=\prod_j\exp\left(-i\frac{\pi}{4}s_j^z\right)$, the Hamiltonian of Eq.~(\ref{eq:Hgamma}) with $J_2=\Gamma_2=0$ is transformed into
\begin{equation}
\mathcal H =\sum_j J_1\left[(1-\gamma)S_j^xS_{j+1}^x +(1+\gamma)S_j^yS_{j+1}^y +S_j^zS_{j+1}^z \right],
\label{eq:XYZoriginal}
\end{equation}
where $S_j=U_z^\dagger s_jU_z$\cite{SaitoH2024PRL,SaitoH2024PRB,Hikihara2026}.
To use Baxter's parametrization\cite{Baxter1972,Baxter1982}, we further rotate the spins by
$U_x=\prod_j\exp\left(-i\frac{\pi}{2}S_j^x\right)$, which permutes the spin axes and yields
\begin{equation}
\mathcal H =\sum_j\left[ \widetilde J_x\widetilde S_j^x\widetilde S_{j+1}^x 
+\widetilde J_y\widetilde S_j^y\widetilde S_{j+1}^y
+\widetilde J_z\widetilde S_j^z\widetilde S_{j+1}^z \right],
\end{equation}
with $\widetilde S_j=U_x^\dagger S_jU_x$ and
\begin{equation}
\widetilde J_x=J_1(1-\gamma),\qquad
\widetilde J_y=J_1,\qquad
\widetilde J_z=J_1(1+\gamma),
\end{equation}
so that $\widetilde J_x<\widetilde J_y<\widetilde J_z$ for $\gamma>0$.
Then we can use Baxter's parametrization to express the ratios between the exchange couplings in terms of elliptic functions:
\begin{equation}
\widetilde J_x/\widetilde J_z=\cn(2\zeta,l),\qquad
\widetilde J_y/\widetilde J_z=\dn(2\zeta,l),
\label{eq:BaxterParam}
\end{equation}
where $l$ is the elliptic modulus.
Using $\sn^2u+\cn^2u=1$ and $\dn^2u+l^2\sn^2u=1$, we obtain $l^2=(2+\gamma)/4$, so that $l\to1/\sqrt2$ for $\gamma\to0$.
Expanding $\cn(2\zeta,l)=(1-\gamma)/(1+\gamma)$ in this limit gives
\begin{equation}
\zeta\simeq\sqrt{\gamma},
\label{eq:zeta_asym}
\end{equation}
which is used to derive the small-$\gamma$ asymptotic form of the one-spinon gap.

The exact spectrum is expressed in terms of a second elliptic modulus $k_1$, defined by
\begin{equation}
K_1'/K_1=\zeta/K_l',
\label{eq:k1def}
\end{equation}
where
$K_1=K(k_1)$, $K_1'=K(k_1')$, $K_l'=K(l')$, and $K(k)$ is the complete elliptic integral of the first kind, with
$k_1'=\sqrt{1-k_1^2}$ and $l'=\sqrt{1-l^2}$.
For $\gamma\to 0$, one has $k_1'\ll1$. Then we use the asymptotic forms\cite{Gradshteyn} of the complete elliptic integrals
\begin{equation}
K(k_1)\simeq\ln (4/k_1') \qquad {\rm and} \qquad K(k_1')\simeq \pi/2,
\end{equation}
to obtain
\begin{equation}
k_1'
\sim4\exp\left[-\frac{\pi K(1/\sqrt2)}{2\sqrt\gamma}\right].
\label{eq:k1prime_asym}
\end{equation}

The exact one-spinon energy \cite{JohnsonJD1973} is given by
\begin{align}
\epsilon_{\rm 1s}(q)=\frac{\widetilde J_z}{2}\sn(2\zeta,l)\frac{K_1}{K_l'}
\sqrt{1-k_1^2\cos^2 q},
\label{eq:1spinon_appendix}
\end{align}
and the one-spinon gap is obtained by evaluating Eq.~(\ref{eq:1spinon_appendix}) at $q=\pi$:
\begin{equation}
\Delta_{1\mathrm{s}} =\frac{\widetilde J_z}{2}\sn(2\zeta,l)\frac{K_1}{K_l'}k_1'.
\end{equation}
Finally, we derive the small-$\gamma$ asymptotic form of the one-spinon gap.
Using $\sn(2\zeta,l)\simeq2\zeta$ and Eq.~(\ref{eq:k1def}),
we obtain
\begin{equation}
\Delta_{1\mathrm{s}}(\gamma)
\simeq2\pi J_1(1+\gamma)\exp\left(-a/\sqrt\gamma\right),
\end{equation}
where $a=\pi K(1/\sqrt2)/2 (=2.912\ldots)$ is a constant.

\end{document}